\documentclass[preprint]{iucr}              
\usepackage{color,xcolor}
\newcommand{\be}{\begin{equation}}
\newcommand{\ee}{\end{equation}}
\RequirePackage{graphicx}
\usepackage{epstopdf}
\usepackage{amsmath}
\usepackage{multirow}
\renewcommand{\vec}{\ensuremath\mathbf}
     \journalcode{A}              

\begin{document}                  



\title{Atomic pair distribution functions (PDFs) from textured polycrystalline samples: fundamentals}
\shorttitle{odfPDF}


\author[a]{Zizhou}{Gong}{}{}
\aff[a]{Department of Physics,
Columbia University, \city{New york, NY 10027}, \country{USA}}
\cauthor[b,c]{Simon J. L.}{Billinge}{}{sb2896@columbia.edu}{}
\aff[b]{Department of Applied Physics and Applied Mathematics, Fu Foundation School of Engineering \& Applied Sciences,
Columbia University, \city{New york, NY 10027}, \country{USA}}
\aff[c]{Condensed Matter Physics and Materials Science Department,
    Brookhaven National Laboratory,
    \city{Upton, NY~11973}, \country{USA}}










\maketitle                        


\begin{abstract}
Equations for the reduced structure function and atomic pair distribution function (PDF) of a 
textured polycrystalline sample are formulated in terms of the orientational
distribution function (ODF) and the structure function from a single crystallite.  This
may be used to determine the sample ODF from experimental data when the structure of the 
reference crystallite is known.
\end{abstract}


\section{\label{sec:Intro}Introduction}

The atomic pair distribution function (PDF) analysis of x-ray, neutron and electron diffraction data is becoming a widely used method for studying the structure of nanomaterials~\cite{laved;jmca17,page;jac11i,stein;cm17}.  Because the methodology does not presume periodic long-range order of the underlying lattice, as is the case in traditional Bragg crystallography, this approach may be extended to nanostructured and disordered systems.  The most commonly applied PDF method starts from the ideal powder approximation, resulting in the 1D PDF, which is simply a histogram of the interatomic distance distributions in the sample~\cite{duxbu;dam16,duxbu;4or16}.  This approximation is a good one for the vast majority of nanocrystalline samples, in part because for nanomaterials the small grain size results in rather good powder averaging.  However, with the recent development of thin-film PDF methods~\cite{jense;ij15}, and the desire to measure nanomaterials in different geometries, the spectre of preferred orientation and crystallographic texture is becoming an issue even for nano-sized grains.

There is no reason, in principle, why the PDF equations cannot be extended to the case of a textured polycrystalline sample.  Here we develop the basic equations for the total scattering structure function and the atomic pair distribution function of a textured polycrystalline sample.

\section{Definition of the polycrystalline structure function, $S_p(\vec{Q})$, and polycrystalline PDF $G_p(\vec{r})$)}

As a start we write down the full 3D structure function~\cite{egami;b;utbp12}, $S(\vec{Q})$, which may be obtained from measured scattering intensities,
\begin{align}
\label{eq:3DS}
S({\vec Q}) &= 1+ \frac{1}{N{\langle f \rangle}^{2}}\sum_{i}\sum_{j > i} f_{i}^{*}(Q) f_{j}(Q)e^{i\vec{Q}\cdot \vec{r}_{ij}}\\
\label{eq:3DS2}
            &= 1+ \frac{1}{N{\langle f \rangle}^{2}}\sum_{i\neq j} f_{i}^{*}(Q) f_{j}(Q)e^{i\vec{Q}\cdot \vec{r}_{ij}},
\end{align}
where $\vec{Q}$ is the scattering vector, $\vec{Q}=\vec{k}_\mathrm{o}-\vec{k}$, where $\vec{k_\mathrm{o}}$ and $\vec{k}$ are the wave vectors of the incident and scattered waves, respectively, and $Q$ is the magnitude of the scattering vector.  $N$ is the number of atoms in the (illuminated part of the) sample and $f_{j}(Q)$ is the atomic form factor of the $i^\mathrm{th}$ atom.  $\langle f \rangle$ is the sample average structure, given by $ \langle f \rangle = \sum_i c_if_i(Q)$.  The sums over $i$ and $j$ run over every atom in the sample in a way that avoids double-counting, where Eq.~\ref{eq:3DS2} serves to define $\sum_{i\neq j}$. Finally,
\begin{equation}
\vec{r}_{ij} = \vec{r}_{j}-\vec{r}_{i},
\end{equation}
is the vector joining atom $i$ and atom $j$, where $\vec{r}_{i}$ is the vector from the origin of the sample reference frame to the $i^\mathrm{th}$ atom.

This function is proportional to the scattered intensity from the sample at $\vec{Q}$.  It depends on $\vec{Q}$ but not on the orientation of the sample.  It is interesting to investigate how this can be possible since the scattered intensity does depend on the sample orientation in a real experiment.  The easiest way to think about this is that we assume a reference frame on the sample (for a crystal this will likely be defined by the crystallographic unit cell) and we vary $\vec{Q}$ to go from the origin of this frame to every voxel in turn in the entire sample-reference frame.  Then we measure the normalized scattered intensity at each point.  In a real experiment it is not possible to place $\vec{Q}$ into every voxel in the sample space without reorienting the sample, so from a practical point of view the scattered intensity is measured for the sample at different orientations in the laboratory reference frame, but each measurement is mapped back to the sample reference frame to yield $S(\vec{Q})$~\cite{ester;pt98}.  To sample the full reciprocal space in an experiment with a large area 2D detector it is possible to rotate the sample around one axis, or two non-collinear axes, that are perpendicular to the incident beam direction. Two axes may be required if there is a missing wedge due incomplete rotation of the sample around one axis.  For brevity we will refer to this as the orthogonal axes rotation (OAR) approach.  To obtain $S(\vec{Q})$ from the measured intensities we make corrections for experimental artefacts, such as various sources of parasitic scattering and multiplicative aberrations such as polarization, absorption and so on.  We also normalize by the incident flux~\cite{egami;b;utbp12}.  In this way, the structure function, $S(\vec{Q})$ may be obtained from any sample with any degree of anisotropy.  For example, when the sample is a single crystal, this function constitutes the 3D crystallographic structure function which may be Fourier transformed to obtain the 3D PDF~\cite{egami;b;utbp12}, which is emerging as a powerful approach for studying diffuse scattering and defects in single crystals~\cite{schau;jac11,egami;b;utbp03,simon;phd14}.  The 3D PDF is given by
\begin{equation}
\label{eq:3DG}
G({\vec r}) = \frac{1}{(2\pi)^3} \int \left[ S({\vec Q})-1 \right] e^{i\vec{Q} \cdot \vec{r}}d{\vec Q}.
\end{equation}

Here we are interested in the particular case where we have a sample that is polycrystalline but has some texture or preferred orientations of crystallites.  The crystallites could be bulk-sized crystals, which is the familiar and widely studied case of a textured polycrystalline sample~\cite{bunge_texture_1982}.  However, as with all PDF studies we do not presume long-range order, and the crystallites could be nanosized in principle.  We seek to understand how scattered intensities from such a sample may be propagated through the Fourier transform to obtain a scientifically relevant real-space pair correlation function, and in principle, how to model that function to obtain information about the texture.

To get a better feel for these 3D structure and PDF functions, we consider two familiar special cases.  If the sample is a single crystal where the average crystal structure solution is desired, two simplifications can speed up data acquisition.  Firstly, the crystallinity results in scattering being confined to small volumes of reciprocal space in the vicinity of reciprocal lattice points.  Once the unit cell is determined, and the orientation of the crystal in a laboratory reference frame (e.g., defined by the UB matrix~\cite{busin;ac67})  it is possible for the single crystal diffractometer to reorient the crystal and detector in such a way as to visit only the volumes of reciprocal space in the vicinity of the reciprocal lattice points and integrate the intensity in those regions, neglecting the large regions of reciprocal space in between.  Furthermore, if the symmetry of the unit cell is known it may be possible to collect a complete dataset by exploring only a subset of the full reciprocal-space volume, although measuring equivalent peaks in different regions of reciprocal-space can help with corrections for experimental artifacts such as sample self-absorption.   If, as is increasingly the case, we are interested also in diffuse scattering in the crystal, then the full reciprocal space volume must be collected, for example, using the OAR approach.  This is becoming highly feasible these days with the use of high-energy x-rays at synchtrotron sources coupled with large area photon counting detectors~\cite{schaub_exploring_2007,schau;jac11,osborn_position-sensitive_1990,welberry_analysis_1998,welberry_analysis_1998-1,proffen_analysis_1997} 
and with neutron diffraction instruments designed for this purpose~\cite{rosenkranz_corelli:_2008,keen_sxd_2006,frost_initial_2010,tamura_current_2012}.

The second special case we consider is a sample that is an isotropic powder with a large number of grains equally sampling all orientations (a good powder average).  In this case there is no need to rotate the sample as the scattering from the powder itself is isotropic. In practice, samples often are spun about an axis perpendicular to the incident beam, especially in synchrotron experiments with small highly parallel beams, to improve the powder statistics and the isotropy of the measured scattering.   In powder measurements, orientational information is lost because the images taken with the sample at different orientations are summed without storing their orientations, and a 1D function, $S(Q)$, is measured. This results in the regular 1D PDF, $G(r)$, when Fourier transformed.  It is assumed in such a case that there is no structural coherence between grains and so the measured structure function is  a superposition of structure functions of identical crystallites taking all possible orientations with equal probability.  Mathematically and experimentally, powder-like data can be obtained from a single crystal by rotating it about an axis parallel to the incident beam at each orientation of the perpendicular rotation.  If a 2D detector is being used, integrating around the Debye-Scherrer rings on the detector is equivalent to rotating the sample about the beam axis, making it somewhat straightforward to obtain a 1D PDF from a 3D PDF dataset for comparison with (or to replace) powder measurements.

Let us now consider the case where the sample is made up of two identical crystallites of the same material that are misoriented with respect to each other, and far enough apart that they are both in the incident beam but beyond the coherence volume of the beam. In other words, we assume that scattering from each crystallite is incoherent and the total observed scattering is just the linear superposition of the scattering from each crystallite (we will assume incoherent scattering between crystallites from hereon out).  We define $\vec{\Omega}$, as the three-vector that contains the Euler angles that define the relative orientation of one crystallite with the other one.  For convenience, and without loss of generality, we assume the sample reference frame is the reference frame of one of the crystallites, which we call the reference crystallite.  If we measured either one of the crystallites as an individual single crystal using the orthogonal axes rotation approach we would get the same single crystal structure function.  However, the measurement is carried out in such a way that the signals from the two crystallites are superposed on the detector.  The crystallite structure function  can be determined if we are able to separate the superposed signals from each crystallite.  For crystalline materials this separation is straightforward and this approach is called polycrystallography and has been developed to a high level~\cite{pouls;b;tdxdm}.


This reasoning is readily extended to the case of $M$ separable diffraction patterns from $M$ crystallites. In this case, as before,  a unique reference frame is defined on a reference crystallite on the sample, which we call the sample reference frame, and we define $\vec{\Omega}_m$ as being the Euler angles that give the orientation of the $m^{\text{th}}$ crystallite with respect to this reference frame.  If $\vec{R}_m$ is the rotation matrix that rotates the sample reference frame onto the $m^{\text{th}}$ crystallite reference frame, we have the following relation
\begin{align}
& \vec{R}_m = \vec{R}(\vec{\Omega}_m), \label{equ:vector-rotation} \\
& \vec{r}^{m}_{ij}= \vec{R}_m\vec{r}_{ij}, \label{eq:rotater}
\end{align}
where $\vec{r}^{m}_{ij}$ refers to the $\vec{r}_{ij}$ interatomic vector of the reference crystallite, but in the $m^{\text{th}}$ crystallite at orientation $\vec{\Omega}_m$.
We can thus write the polycrystalline sample-structure function $S_p(\vec{Q})$, as
\begin{align}
S_p(\vec{Q}) &= 1+ \frac{1}{N{\langle f \rangle}^{2}}\sum_{m}\sum_{i\neq j}  f_{i}^{*} f_{j}e^{i\vec{Q}\cdot \vec{r}^{m}_{ij}}\\
&= 1+ \frac{1}{N{\langle f \rangle}^{2}}\sum_{m}\sum_{i\neq j}  f_{i}^{*} f_{j}e^{i\vec{Q}\cdot (\vec{R}_m\vec{r}_{ij})\label{eq:3DSp}
}
\end{align}
where for notational simplicity we have dropped the explicit $Q$-dependence of the atomic form factors. The double-sum over $i$ and $j$ is now a sum over the interatomic vectors between just the atoms in the reference crystallite and the sum over $m$ is a sum over all the (assumed to be) identical but misoriented crystallites. In Equation~\ref{eq:3DSp} the signal for the polycrystalline sample is built up by rotating the reference crystallite to the orientation of each crystallite in the sample.

We now turn to a polycrystalline sample with a large number of crystallites where the scattering from the individual crystallites are no longer separable, but the sample is still not isotropic: a textured powder.  The patterns from the individual grains strongly overlap and multiple crystallites contribute to each region (voxel) of reciprocal space defined by the $\vec{Q}$ resolution of our measurement.  In this case we would like to convert Eq.~\ref{eq:3DSp} to a continuous function. We define a volume element $d\vec{\Omega}$ in the Euler angle space that runs from $(\theta,\phi,\xi)$ to $(\theta+d\theta,\phi+d\phi,\xi+d\xi)$.  We can then define the number of crystallites in the beam that have an orientation such that their Euler angles place it in that volume element of angle-space as $N(\vec{\Omega})$.

Now, returning to Eq.~\ref{eq:3DSp}, we would like to rewrite this equation in terms of a sum over all orientation directions rather than a sum over $m$.   Denoting the total number of crystallites within the sample as $n_0$, the total number of atoms in the sample, $N$, is then given by
\begin{equation}
N = N^{'} \cdot n_{0},
\end{equation}
where $N^{'}$ is the number of atoms in the reference crystallite.  The sum over $m$ then becomes
\begin{equation}
\sum_{m}1 = \sum_{lnp} N(\vec{\Omega}_{lnp}) = n_{0},
\end{equation}
where $lnp$ label voxels in the orientation space and the sum runs over all voxels in the orientation space. Furthermore, since the crystallite with the same orientation $\vec{\Omega}$ gives the same contribution to $S_p(\vec{Q})$, we can rewrite $S_p(\vec{Q})$ as the summation over different crystallite orientations, weighted by the number of crystallites with that orientation:
\begin{align}
S_{p}(\vec{Q})
& = 1+ \frac{1}{N{\langle f \rangle}^{2}}\sum_{m}\sum_{i\neq j}  f_{i}^{*} f_{j}e^{i\vec{Q}\cdot (\vec{R}_m\vec{r}_{ij})} \\
& = 1 + \frac{1}{N{\langle f \rangle}^{2}} \sum_{lnp}  N(\vec{\Omega}_{lnp}) \sum_{i\neq j}  f_{i}^{*} f_{j} e^{i\vec{Q}\cdot \left(\vec{R}_{\vec{\Omega}_{lnp}}\vec{r}_{ij}\right)} \label{eq:spsumtoint}.
\end{align}

For crystallites oriented quasi-continuously in every orientation we rewrite this equation with an integral, using $N(\vec{\Omega})=n(\vec{\Omega})d\vec{\Omega}$,
\begin{align}
S_{p}(\vec{Q})
& = 1 + \frac{1}{N {\langle f \rangle}^{2}} \int n(\vec{\Omega})  \sum_{i \neq j}  f_{i}^{*} f_{j} e^{i\vec{Q}\cdot \left(\vec{R}_{\vec{\Omega}}\vec{r}_{ij}\right)} d\vec{\Omega}\\
 \label{eq:spdiscOmega}
& = 1 + \frac{1}{N^{'}{\langle f \rangle}^{2}}  \int  \frac{n(\vec{\Omega})}{n_{0}}  \sum_{i\neq j}  f_{i}^{*} f_{j} e^{i\vec{Q}\cdot (\vec{R}_{\vec{\Omega}}\vec{r}_{ij})} d\vec{\Omega} \\
& = 1 + \frac{1}{N^{'}{\langle f \rangle}^{2}}  \int  D(\vec{\Omega})  \sum_{i\neq j}  f_{i}^{*} f_{j} e^{i\vec{Q}\cdot (\vec{R}_{\vec{\Omega}}\vec{r}_{ij})} d\vec{\Omega}, \label{eq:spqo0final}
\end{align}
where we introduce the orientation distribution function (ODF), $D(\vec{\Omega}) = n(\vec{\Omega})/n_0$.
In practice, $S_{p}(\vec{Q})$ can be evaluated by exchanging the order of the summation over $i$ and $j$ with the integration over $\vec{\Omega}$ and evaluate the integral involving the ODF and the complex exponential factor.

Here in Equation~\ref{eq:spqo0final} the function $D(\vec{\Omega})$ has the meaning of the fraction of the crystallites with orientation $\vec{\Omega}$ among all crystallites in the sample. It is a sample-dependent property, not depending explicitly on sample orientation, and is expressed in the sample reference frame. Since by definition the ODF is a probability density, it has the normalization property:
\begin{align}
\label{equ:odf_normalization}
\int D(\vec{\Omega})d\vec{\Omega} = 1.
\end{align}

In the special case that we have considered here, the sample is assumed to consist of many identical crystallites that all have the same structure function, $S^{'}(\vec{Q})$, of the reference crystallite but are oriented with respect to that crystallite by $\vec{\Omega}$.  To capture this we introduce a generalized structure function, for the ``misoriented" crystallites,
\begin{equation}
S^{'}(\vec{Q},\vec{\Omega})  =  1 +  \frac{1}{N^{'}{\langle f \rangle}^{2}}  \sum_{i\neq j}   f_{i}^{*} f_{j} e^{i\vec{Q}\cdot (\vec{R}_{\vec{\Omega}}\vec{r}_{ij})}.
\end{equation}

We can then rewrite the polycrystalline sample structure function in terms of $S^{'}(\vec{Q},\vec{\Omega})$, taking advantage of the normalization property of the ODF in Eq.~\ref{equ:odf_normalization}. First, we change the order of integration and summing,
\begin{align}
S_{p}(\vec{Q})
& = 1 + \frac{1}{N^{'}{\langle f \rangle}^{2}}  \int  D(\vec{\Omega}) \sum_{i\neq j}  f_{i}^{*} f_{j} e^{i\vec{Q}\cdot (\vec{R}_{\vec{\Omega}}\vec{r}_{ij})} d\vec{\Omega} \\
& = 1 + \frac{1}{N^{'}{\langle f \rangle}^{2}}  \sum_{i\neq j}  f_{i}^{*} f_{j} \int  D(\vec{\Omega}) e^{i\vec{Q}\cdot (\vec{R}_{\vec{\Omega}}\vec{r}_{ij})} d\vec{\Omega} \\
& = 1 + \frac{1}{N^{'}{\langle f \rangle}^{2}}  \sum_{i\neq j}  f_{i}^{*} f_{j} I_{ij}^D(\vec{Q})
\end{align}
which serves to define the integral
\begin{equation}
I_{ij}^D(\vec{Q}) = \int  D(\vec{\Omega}) e^{i\vec{Q}\cdot (\vec{R}_{\vec{\Omega}}\vec{r}_{ij})}d\vec{\Omega} \label{eq:ourintegral}.
\end{equation}
Now, taking advantage of the normalization property of our ODF we can write
\begin{align}
S_{p}(\vec{Q}) & =   \int  D(\vec{\Omega}) d\vec{\Omega}  +   \int  D(\vec{\Omega})  \frac{1}{N^{'}{\langle f \rangle}^{2}}\sum_{i \neq j}  f_{i}^{*} f_{j} e^{i\vec{Q}\cdot (\vec{R}_{\vec{\Omega}}\vec{r}_{ij})} d\vec{\Omega} \\
& = \int D(\vec{\Omega}) S^{'}(\vec{Q},\vec{\Omega}) d\vec{\Omega}. \label{eq:sqsprimeq}
\end{align}
Equation~\ref{eq:sqsprimeq} expresses the structure function of the sample as an orientational distribution weighted arithmetic average of the structure function of the reference crystallite.

We note that Equation~\ref{eq:sqsprimeq} for the case of discrete and separable crystallites may also be rewritten in this way as
\begin{align}
S_p(\vec{Q})
& = 1+ \frac{1}{N{\langle f \rangle}^{2}}\sum_{m}\sum_{i\neq j}  f_{i}^{*} f_{j}e^{i\vec{Q}\cdot (\vec{R}_m\vec{r}_{ij})} \\
& = \sum_{m} \frac{1}{n_0}  + \sum_{m} \frac{1}{N^{'}n_0 {\langle f \rangle}^{2}}\sum_{i\neq j}  f_{i}^{*} f_{j}e^{i\vec{Q}\cdot (\vec{R}_m\vec{r}_{ij})} \\
& = \frac{1}{n_0}\sum_{m} \left( 1+ \frac{1}{N^{'}{\langle f \rangle}^{2}}\sum_{i\neq j}  f_{i}^{*} f_{j}e^{i\vec{Q}\cdot (\vec{R}_m\vec{r}_{ij})} \right) \\
& =\frac{1}{n_0} \sum_m S^{'}(\vec{Q},\vec{\Omega}_m),  \label{eq:spsprimedisc}
\end{align}
Equations~\ref{eq:spsprimedisc} and~\ref{eq:sqsprimeq} hold when the approximation that the sample is made up of multiple identical crystallites, or nanoparticles, that have different orientations.

We now consider how this propagates through the Fourier transform to yield a textured polycrystalline pair correlation function, $G_{p}({\vec r})$,
\begin{align}
G_{p}({\vec r})
& = \frac{1}{(2\pi)^3} \int \left[ S_{p}({\vec Q})-1 \right] e^{-i\vec{Q} \cdot \vec{r}}d{\vec Q} \\
& =  \int \left[\frac{1}{(2\pi)^3} \int D(\vec{\Omega}) \left( S^{'}({\vec Q},\vec{\Omega})-1 \right) d\vec{\Omega} \right]  e^{-i\vec{Q \cdot r}} d{\vec Q} \\
& = \int D(\vec{\Omega}) \left[\frac{1}{(2\pi)^3} \int   \left( S^{'}(\vec{Q},\vec{\Omega})-1  \right)  e^{-i\vec{Q \cdot r}} d{\vec Q} \right]  d\vec{\Omega} \\
&  = \int D(\vec{\Omega}) G^{'}({\vec r},\vec{\Omega} )  d\vec{\Omega} .
\end{align}
Where $G^{'}({\vec r},\vec{\Omega})$ is, following the definition of PDF in Equation~\ref{eq:3DG}, the 3D PDF from of the reference crystallite but with orientation $\vec{\Omega}$, with respect to the sample reference frame, expressed as
\begin{equation}
G^{'}({\vec r},\vec{\Omega}) =  \frac{1}{(2\pi)^3}  \int \left( S^{'}(\vec{Q},\vec{\Omega})-1 \right)  e^{-i\vec{Q \cdot r}} d{\vec Q}.
\end{equation}

These equations serve to define the real and reciprocal-space representations of textured polycrystalline samples.  In general, $S_p(\vec{Q})$ may be measured in the same way as we measure the 3D PDF of a single-crystal, for example, using the OAR method with x-rays or in a neutron single-crystal experiment.  If, as is often the case, we know $S^{'}(\vec{Q})$, the structure function of the reference crystallite, we can build up the polycrystalline intensity at each $\vec{Q}$ by rotating $S^{'}(\vec{Q})$ to all angles and adding the contribution.  We note that the derivation did not assume crystallinity of the sample, and so it is equally applicable to polycrystalline textured nanoparticle samples and non-isotropic amorphous samples, provided that in these samples the approximation that the local clusters are all equivalent to each other apart from their orientation.  A similar approach could be carried out directly in real-space to determine $G_{p}({\vec r})$.  If it is desired to determine the ODF, this approach may be implemented in a regression loop.


\section{Acknowledgements}
This work was supported by U.S. Department of Energy, Office of Science, Office of Basic Energy Sciences (DOE-BES) under contract No. DE-SC00112704.


\end{document}